\documentclass[twocolumn,preprintnumbers,superscriptaddress,nofootinbib,aps,prd,floatfix]{revtex4}

\usepackage{amsmath,amssymb}
\usepackage{graphicx}
\usepackage{epstopdf} 
\usepackage{slashed}
\usepackage{subfigure}
\usepackage{color}
\usepackage{multirow,footmisc}
\usepackage{hyperref}
\usepackage{enumitem}
\usepackage{xspace}
\usepackage{lipsum}

\usepackage{footmisc,ulem}

\usepackage{array}
\newcolumntype{L}[1]{>{\raggedright\let\newline\\\arraybackslash\hspace{0pt}}m{#1}}
\newcolumntype{C}[1]{>{\centering\let\newline\\\arraybackslash\hspace{0pt}}m{#1}}
\newcolumntype{R}[1]{>{\raggedleft\let\newline\\\arraybackslash\hspace{0pt}}m{#1}}

\newcommand{\be}{\begin{eqnarray*}}
\newcommand{\ee}{\end{eqnarray*}}

\newcommand{\bee}{\begin{eqnarray}}
\newcommand{\eee}{\end{eqnarray}}
\newcommand{\beeq}{\begin{equation}}
\newcommand{\eeeq}{\end{equation}}

\newcommand{\pt}{\ensuremath{p_{\mathrm{T}}}\xspace}
\newcommand{\ptH}{\ensuremath{p_{\mathrm{T}}^{H}}\xspace}

\begin{document}

%%%%%%%%%%%%%%%%%%%%%%%%%%%%%%%%%%%%%%%%%%%%%%%%%%%%%%%%%%
\title{Higgs characterisation in the presence of theoretical uncertainties and invisible decays}
%%%%%%%%%%%%%%%%%%%%%%%%%%%%%%%%%%%%%%%%%%%%%%%%%%%%%%%%%%
\begin{abstract}
While the Higgs characterisation programme is well underway, direct signs for new physics beyond the Standard Model remain elusive. Performing a fit of fully differential Higgs production cross sections at the LHC to a subset of Higgs-relevant effective operators, we discuss the extent to which theoretical uncertainties can limit the sensitivity in such a new physics search programme. Extending the dimension-6 Higgs Effective Field Theory framework by introducing new light degrees of freedom that can contribute to an invisible (or undetectable) Higgs decay width $h\to \phi\phi$, we show how differential coupling fits can disentangle effects from non-Standard Model couplings and an invisible decay width, as present in many new physics scenarios, such as Higgs-portal dark matter. Including the so-called off-shell measurement that has been advocated as a sensitive determination of the Higgs width in the $\kappa$ framework, we show explicitly that this method does not provide complementary sensitivity for scale-separated new physics $\Lambda\gg m_h \gg m_\phi$, which is favoured in beyond the Standard Model scenarios that relate astrophysics and collider phenomenology in light of non-observation of new physics during run 1 of the LHC.
\end{abstract}
%%%%%%%%%%%%%%%%%%%%%%%%%%%%%%%%%%%%%%%%%%%%%%%%%%%%%%%%%%
%
%
\author{Christoph Englert} \email{christoph.englert@glasgow.ac.uk}
\affiliation{SUPA, School of Physics and Astronomy, University of
  Glasgow,\\Glasgow G12 8QQ, UK\\[0.2cm]}
\author{Roman Kogler} \email{roman.kogler@uni-hamburg.de}
\affiliation{Institut f\"ur Experimentalphysik, Universit\"at Hamburg,
  22761 Hamburg, Germany\\[0.2cm]}
\author{Holger Schulz} \email{holger.schulz@durham.ac.uk}
\affiliation{Institute for Particle Physics Phenomenology, Department
  of Physics,\\Durham University, Durham DH1 3LE, UK\\[0.2cm]}
\author{Michael Spannowsky} \email{michael.spannowsky@durham.ac.uk}
\affiliation{Institute for Particle Physics Phenomenology, Department
  of Physics,\\Durham University, Durham DH1 3LE, UK\\[0.2cm]}

\pacs{}
\preprint{IPPP/17/60, DCPT/17/120}

\maketitle

%%%%%%%%%%%%%%%%%%%%%%%%%%%%%%%%%%%%%%%%
\section{Introduction}
\label{sec:intro}
%%%%%%%%%%%%%%%%%%%%%%%%%%%%%%%%%%%%%%%%
The discovery of the Higgs boson~\cite{Aad:2012tfa,Chatrchyan:2012ufa}, while marking a milestone in clarifying the mechanism of spontaneous symmetry breaking, has left us with more open questions than solutions. On the one hand, at this stage of the LHC phenomenology programme the nature of the TeV scale is only poorly understood; electroweak symmetry breaking in the Standard Model (SM) is ad-hoc. On the other hand, the unsuccessful search for new interactions beyond the SM (BSM) highlights the fact that all phenomenological aspects of the TeV scale seem to be well-described by the SM paradigm. 

The continued success story of the SM, which has also prevailed for the first 13 TeV measurements, has had a sobering effect on the potentially too optimistic expectations of a new physics discovery early on in the LHC era. However, the lack of concrete evidence for new interactions has resulted in the formulation of a Higgs boson characterisation programme in much greater detail than previously anticipated. The particle physics community as a whole is moving towards understanding the SM as a low energy effective field theory (EFT)~\cite{Passarino:2012cb,Jenkins:2013sda,Jenkins:2013wua,Alonso:2013hga,Ghezzi:2015vva,Berthier:2015gja,Hartmann:2015aia} (see also e.g.\ Ref.~\cite{Aad:2015tna} for experimental results). Where an EFT description is not sufficient anymore to accommodate new phenomenological aspects, ``simplified models''\footnote{See for example Ref.~\cite{Alves:2011wf} for and overview.} have entered the stage.\footnote{Simplified models can be understood in terms of EFTs where the SM particle content has been extended by light degrees of freedom.} These approximations to motivated UV completions are well studied in the context of dark matter searches at colliders and beyond, and they have quickly become the lingua franca to report results~\cite{Abercrombie:2015wmb} (see also Refs.~\cite{Aaboud:2017uak,Sirunyan:2017hci} for recent representative searches performed by the ATLAS and CMS Collaborations).

The special role of the Higgs boson as a direct manifestation of electroweak symmetry breaking can be the harbinger of BSM effects, not only related to anomalous interactions with known matter. The possibility to form SM singlet operators $\sim H^\dagger H$ at a renormalisable level opens the possibility to interpret the electroweak scale within the broader realm of established phenomena beyond the SM, such as dark matter~\cite{Djouadi:2011aa,Lebedev:2011iq}, new approaches to a natural TeV scale, leptogenesis~\cite{Khoze:2016zfi}, or even dark energy~\cite{Bertolami:2007wb}. In most of these scenarios, the extension of the SM by a renormalisable operator 
\begin{equation}
\label{eq:portal}
	{\cal L}_{BSM}= {\cal L}_{SM} + {\cal{O}}_{\text{hid}} |H|^2
\end{equation}
that may or may not be charged under one or multiple hidden sectors gives enough freedom to at least partially relate the electroweak scale to BSM physics~\cite{Schabinger:2005ei, Chacko:2005pe, Strassler:2006im, Englert:2011yb, Englert:2013gz}. Often, addressing these phenomena requires the new hidden degrees of freedom to be light compared to the electroweak scale, ${\cal{O}}_{\text{hid}}\sim \phi^2$, with $m_\phi \ll v$ (we will denote the mass of the new SM-singlet scalar of Eq.~\eqref{eq:portal} with $m_\phi$ in the following).

In such scenarios, interactions of type Eq.~\eqref{eq:portal} can introduce new and invisible decay channels $h\to \phi\phi$ of the physical Higgs which can be tackled by adapted search strategies. Such an invisible branching ratio, irrespective of how it is measured, has far reaching consequences for the relation between the Higgs boson and dark matter.

One strategy that has been motivated recently to constrain (or eventually measure) the invisible Higgs decay width is through the so-called off-shell measurements in $gg\to ZZ$~\cite{Kauer:2012hd,Caola:2013yja,Khachatryan:2014iha,Aad:2015xua,Khachatryan:2016ctc}. Focussing on the intrinsic dependence of this process on new heavy particles in the loop~\cite{Englert:2014aca,Azatov:2014jga,Buschmann:2014twa,Buschmann:2014sia,Cacciapaglia:2014rla,Ghezzi:2014qpa,Englert:2014ffa}, one can give an interpretation of this measurement within the dimension-6 EFT framework~\cite{Buchmuller:1985jz,Hagiwara:1986vm,Grzadkowski:2010es} along the lines of Eq.~\eqref{eq:portal}
\begin{equation}
	\label{eq:modhiggs}
	\Gamma_h=\Gamma_h^{\text{SM}} + \Gamma_h^{\text{D6}} + \Gamma_h^{\text{inv}}\,.
\end{equation}
From a UV perspective such a model has several scales with $m_\phi \ll v \ll \Lambda$, where $\Lambda$ refers to the scale of new (unspecified) interactions. A well-known and well-studied UV completion of this simplified model could be the NMSSM (see e.g.\ Ref.~\cite{Butter:2015fqa}). It is also important to note that modifying the Higgs width in Eq.~\eqref{eq:modhiggs} amounts to setting constraints on all experimentally non-resolvable decay widths, e.g.\ first generation quarks or gluons with (or without) the condition $\Gamma_h^{\text{inv}}=0$.

In this paper we extend differential Higgs fits~\cite{Englert:2015hrx} with new decay channels as outlined above and study the extent to which limits can be set through a differential measurement of Higgs distributions. Our results are concerned with a target luminosity of $3\,\text{ab}^{-1}$ at a centre-of-mass energy of 14 TeV. 

Central to such a projection is the theoretical accuracy that will be available at this stage of the LHC programme (see similar discussions in the context of top sector fits~\cite{Englert:2016aei,Englert:2017dev}). Any related assumption that is made in this moment in time is guesswork and can be criticised as either too optimistic or too pessimistic. However, tracing the influence of theoretical uncertainties down to the impact on the total Higgs width allows a clear avenue to discuss the influence of differential as well as total inclusive theoretical uncertainties as we will show below. The merit of studying effective theories of the form of Eq.~\eqref{eq:portal} is hence two-fold: we do not only discuss the quantitative impact of BSM physics on the Higgs sector related to light degrees of freedom, but we also point out directions how BSM searches in the context of Higgs measurements can be systematically improved with regards to maximum phenomenological impact.

This paper is organised as follows. In Sec.~\ref{sec:diffhiggs} we briefly recapitulate our fit procedure before we discuss how Higgs \pt-dependent uncertainties limit the sensitivity reach to new generic physics in the Higgs sector both qualitatively and quantitatively. In Sec.~\ref{sec:conshiggs} we use these results to discuss the impact of such measurements on the total Higgs width in the scenario given by Eq.~\eqref{eq:modhiggs}. Section~\ref{sec:offshell} is devoted to the inclusion of the off-shell region of $gg\to ZZ$, where we discuss in detail the additional information that is gained by this measurement for the EFT+singlet extension. We summarise and conclude in Sec.~\ref{sec:conc}.

%%%%%%%%%%%%%%%%%%%%
\section{Differential Higgs fits and uncertainties}
\label{sec:diffhiggs}

In the following we will focus on the operators reported in~\cite{Giudice:2007fh}

\begin{widetext}
\begin{equation}
\label{eq:silh}
\begin{split} {\cal L}_{\text{SILH}} = &\frac{\bar
c_H}{2v^2}\partial^\mu \left( H^\dagger H \right) \partial_\mu \left(
H^\dagger H \right)
    + \frac{\bar c_T}{2v^2}\left (H^\dagger {\overleftrightarrow {
D^\mu}} H
    \right)  
    \left(   H^\dagger{\overleftrightarrow D}_\mu H\right) 
    - \frac{\bar c_6\lambda}{v^2}\left( H^\dagger H \right)^3  
    \\ 
    &+ \left( \frac{\bar c_{u,i} y_{u,i}}{v^2}H^\dagger H  {\bar
        u}_L^{(i)} H^c u^{(i)}_R +{\text{h.c.}}\right)  
  + \left( \frac{\bar c_{d,i} y_{d,i}}{v^2}H^\dagger H {\bar d}_L^{(i)} H d^{(i)}_R+{\text{h.c.}}\right) 
    \\
    &+\frac{i\bar c_Wg}{2m_W^2}\left( H^\dagger  \sigma^i \overleftrightarrow
      {D^\mu} H \right )( D^\nu  W_{\mu \nu})^i  
    +\frac{i\bar c_Bg'}{2m_W^2}\left( H^\dagger  \overleftrightarrow {D^\mu}
      H \right )( \partial^\nu  B_{\mu \nu})  
    \\ 
    & +\frac{i\bar c_{HW} g}{m_W^2}
    (D^\mu H)^\dagger \sigma^i(D^\nu H)W_{\mu \nu}^i
    +\frac{i\bar c_{HB}g'}{m_W^2}
    (D^\mu H)^\dagger (D^\nu H)B_{\mu \nu}
    \\
    &   +\frac{\bar c_\gamma {g'}^2}{m_W^2}H^\dagger H
    B_{\mu\nu}B^{\mu\nu} 
    +\frac{\bar c_g g_S^2}{m_W^2}H^\dagger H
    G_{\mu\nu}^a G^{a\mu\nu}\,,
\end{split} 
\end{equation}
\end{widetext}
and we consider operators $\sim \bar{c}_{u,i},\bar{c}_{d,i}$ for the third generation of fermions only and identify $\bar c_T=0$ as well as $\bar c_W + \bar c_B =0$ as the usual approach to reflect LEP constraints (see Ref.~\cite{Berthier:2015gja} for a more dedicated analysis). As we do not consider precision studies~\cite{Gorbahn:2016uoy,Degrassi:2016wml,Bizon:2016wgr,Degrassi:2017ucl,Kribs:2017znd} or di-Higgs constraints~\cite{Goertz:2014qta, Azatov:2015oxa, DiVita:2017eyz}, we neglect $\bar{c}_6$. 

This list of operators is not exhausting all
possibilities~\cite{Contino:2013kra}, but it provides a sufficiently general,
while computationally manageable, theoretical framework to gauge the
sensitivity to BSM effects in the Higgs sector. The details of our fit setup
have been provided in Ref.~\cite{Englert:2015hrx} are only briefly 
    summarised here. Cross sections are generated in a linearised
    approach ($\sim c_i$ in Eq.~\eqref{eq:silh}) using the phase-space
    integration of {\sc{Vbfnlo}}~\cite{Arnold:2012xn} that is interfaced with
    {\sc{FeynArts}}, {\sc{FormCalc}},
    {\sc{LoopTools}}~\cite{Hahn:1998yk,Hahn:2000kx}, and
    {\sc{FeynRules}}~\cite{Christensen:2008py,Alloul:2013bka,Alloul:2013jea}.
    Branching ratios are calculated with {\sc{eHdecay}}~\cite{Contino:2014aaa}.
    Concretely this means we keep dimension-6 amplitude ${\cal{M}}_{d=6}$
    separate from the SM contribution ${\cal{M}}_{\text{SM}}$ and only consider
    the interference of the dimension-6 amplitudes with the SM. The histograms containing the production cross sections 
    and branching ratios are interpolated with
    {\sc{Professor}}~\cite{Buckley:2009bj}, which provides an analytical
    parametrisation that is then fast and accurate enough for the statistical evaluation
    based on the {\sc{Gfitter}} framework~\cite{Flacher:2008zq, Baak:2011ze,
    Baak:2012kk, Baak:2014ora}. 
The accuracy of the {\sc{Professor}} parameterisation is below $0.1\%$ when using polynomials of order five.

Our motivation to work with strictly linearised (pseudo-)observables relates to a very general understanding of validity of the EFT approach. The perturbative means we use to calculate cross sections are challenged when the interference contribution becomes comparable to the SM part: $2\hbox{Re}({\cal{M}}_{d=6} {\cal{M}}_{\text{SM}}) \sim |{\cal{M}}_{\text{SM}}|^2$, which is then typically accompanied by negative cross sections in differential bins. This problem is familiar from higher order QCD calculations, where negative bin contents signal the requirement for further corrections. In our case, the appearance of negative cross sections in integrated bins signallises the ultimate breakdown of perturbative techniques for a given value of $c_i$ in Eq.~\eqref{eq:silh}. Given that we linearise the contributions from new physics in every bin, this is symmetrically reflected also in cross section excesses. Note that we do not impose any constraints from unitarity arguments and our constraints can be interpreted in strongly interacting scenarios, however, within the limitations of our fundamentally perturbative approach to calculating pseudo-observables.

For the sensitivity studies we consider the production modes $pp\to H$, $pp \to H+{\text{j}}$, $pp
\to t\bar t H$, $p p \to WH$, $p p \to ZH$ and $p p\to H+2{\text{j}}$
(via gluon fusion and weak boson fusion) and assume inclusive signal strength measurements 
and measurements of fully differential Higgs transverse momentum distributions for a HL-LHC scenario with $3\,\text{ab}^{-1}$. 
Assuming realistic losses due to experimental acceptances and efficiencies, this results in a total of 46 signal strength measurements and 117 measured bins from differential Higgs transverse momentum distributions~\cite{Englert:2015hrx}. The experimental systematic uncertainties are obtained through a luminosity scaling of the present available uncertainties. This results in improved systematic uncertainties in the low energy regions and uncertainties mostly dominated by statistical components in the tails of the distributions. Additionally, for our studies including the measurement of $gg\to ZZ$, we include 18 bins in the invariant mass of the $ZZ$ system with $m_{ZZ}>330~\mathrm{GeV}$, where the precision of these pseudo-data is statistically limited. In a given production and decay channel, experimental uncertainties are included as correlated uncertainties in our setup.

%%%%%%%%%%%%%%%%%%%%
\begin{figure*}[!t]
	\includegraphics[width=0.47\textwidth]{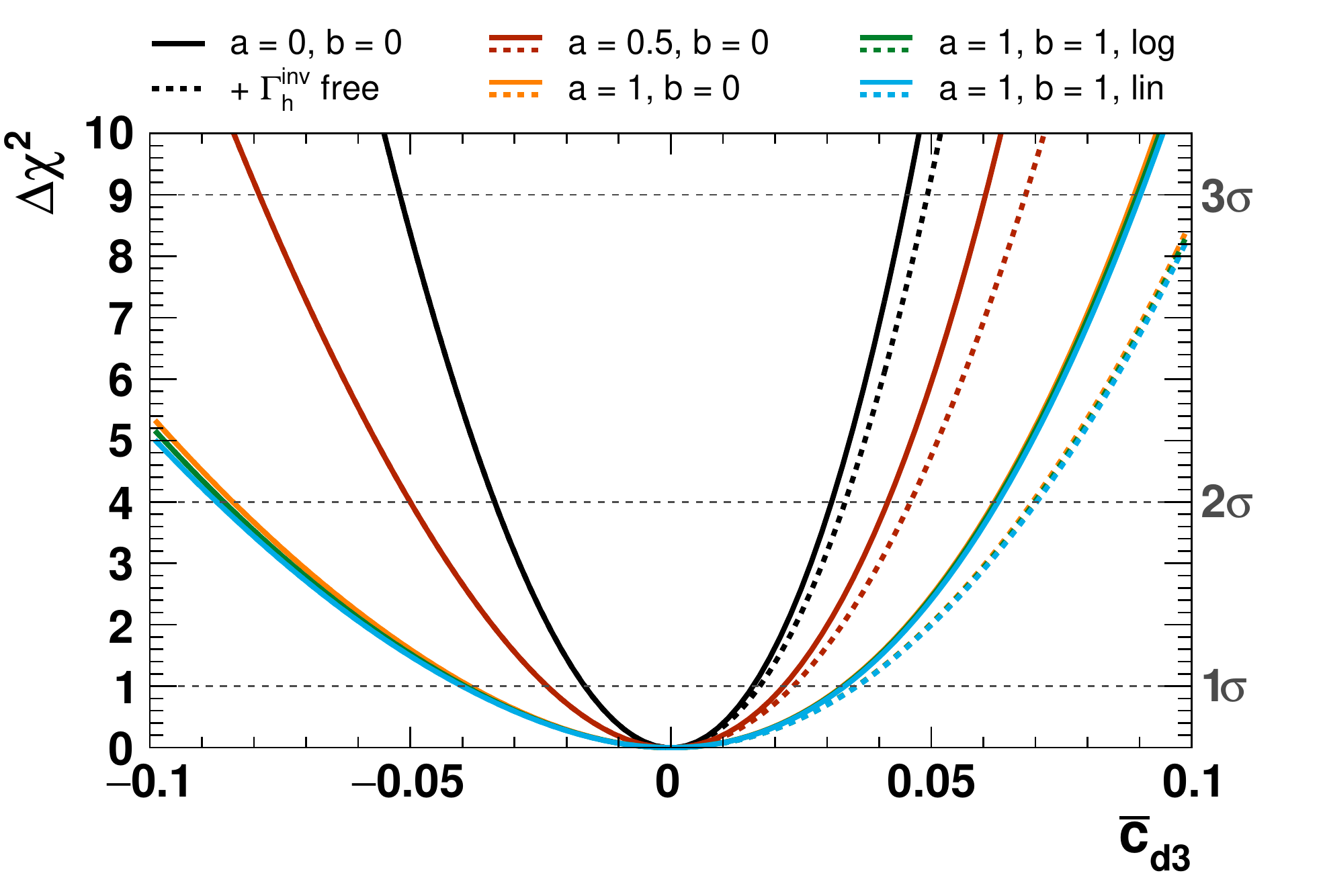}
	\hfill
	\includegraphics[width=0.47\textwidth]{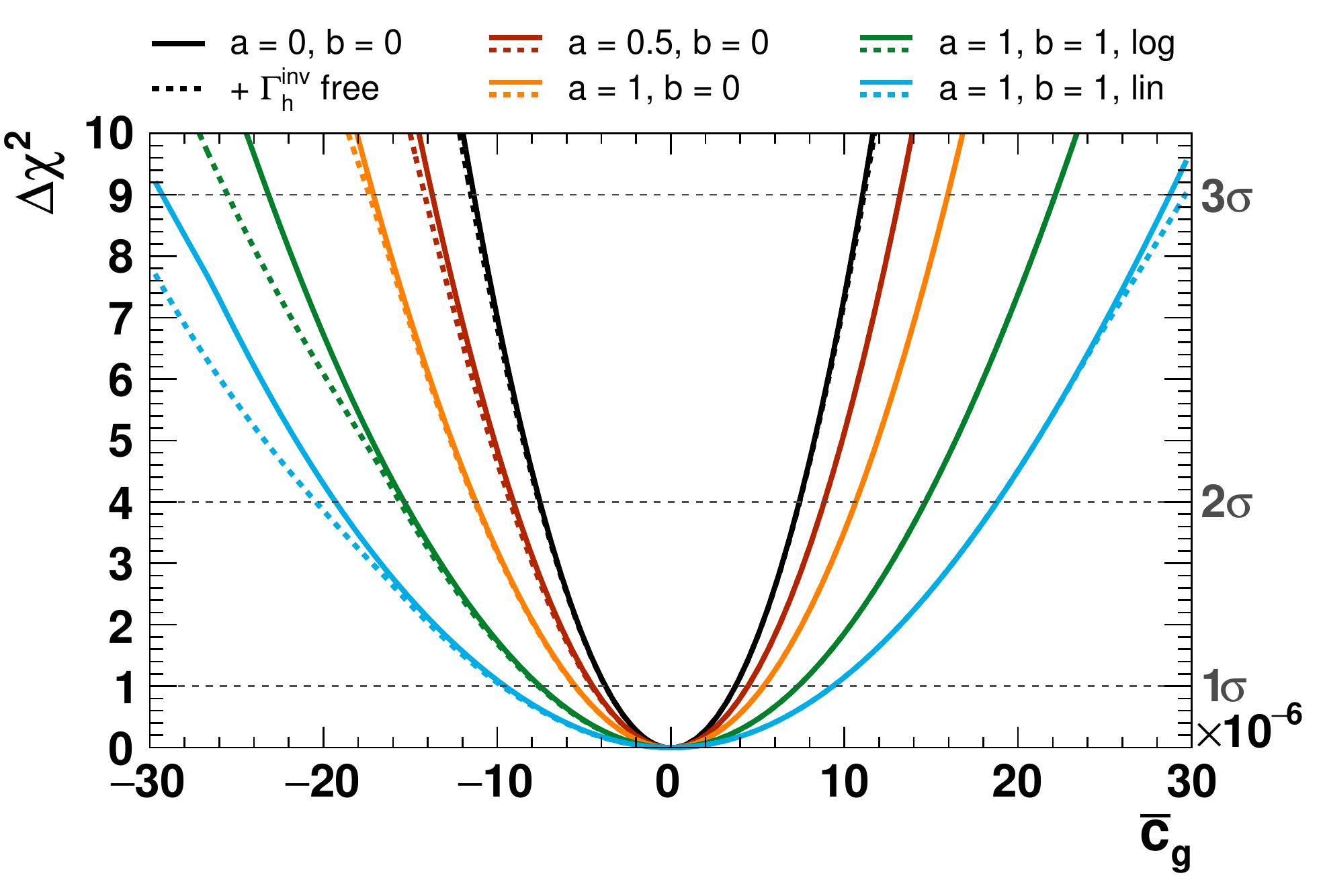} \\
	\includegraphics[width=0.47\textwidth]{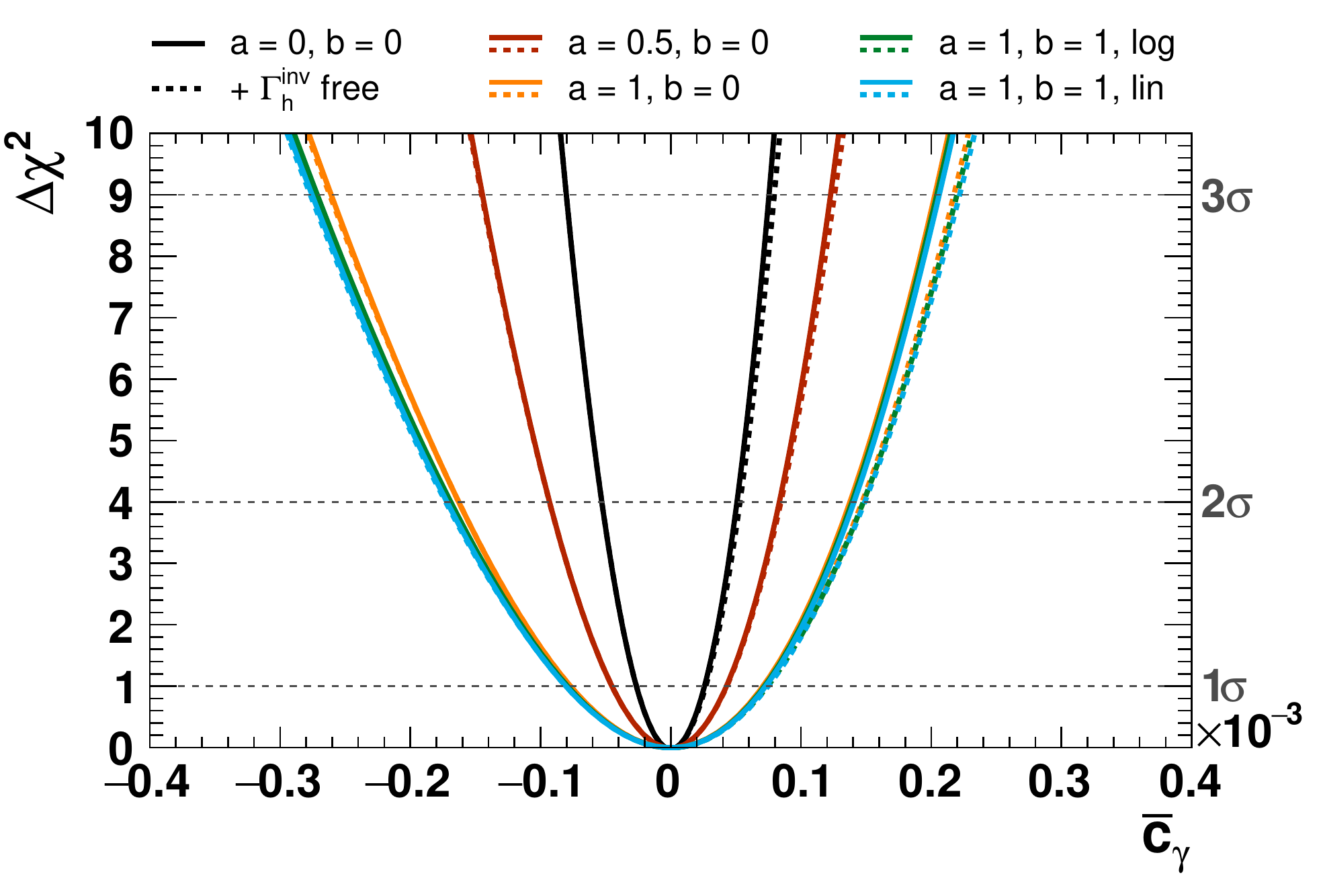}
	\hfill
	\includegraphics[width=0.47\textwidth]{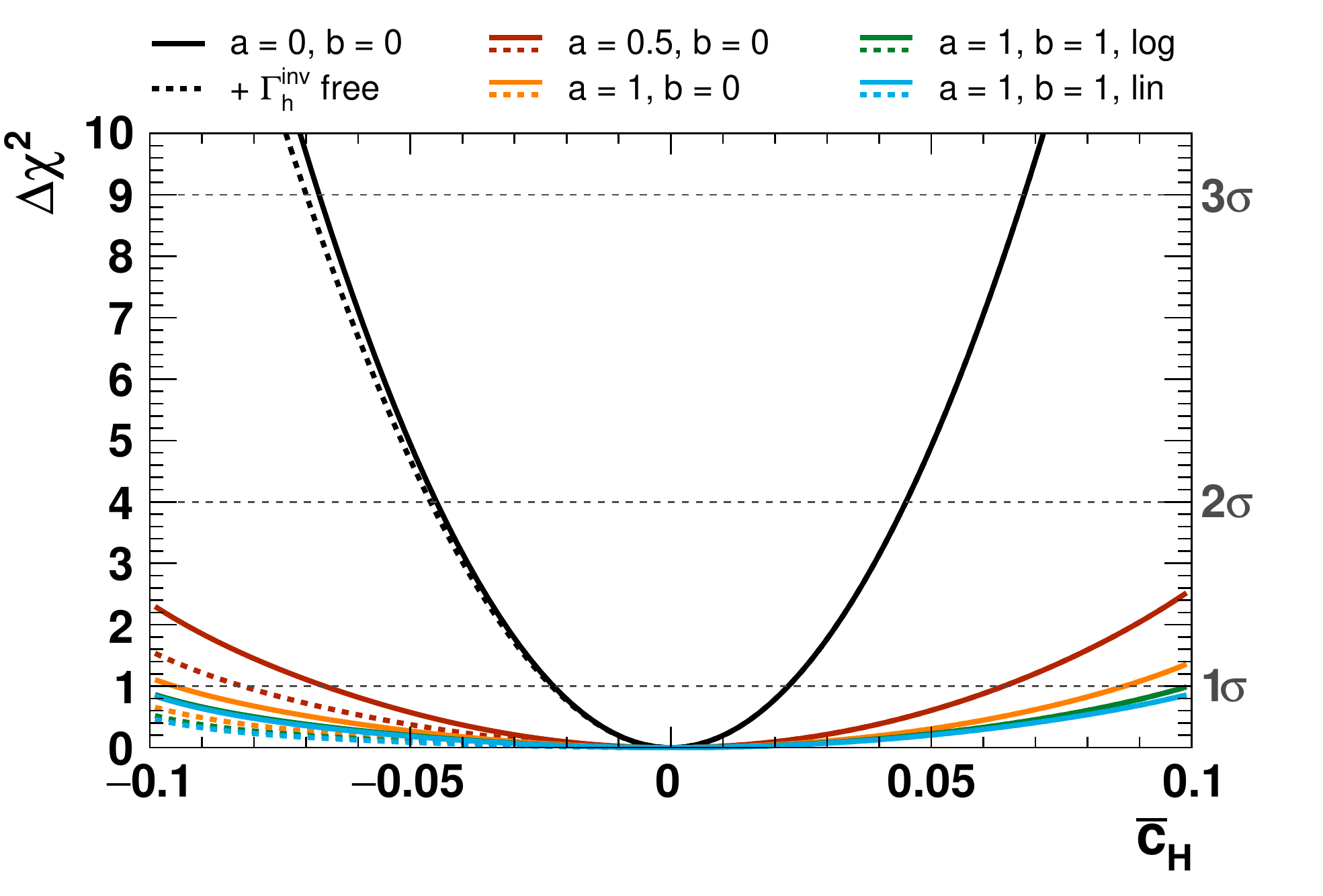} \\
	\includegraphics[width=0.47\textwidth]{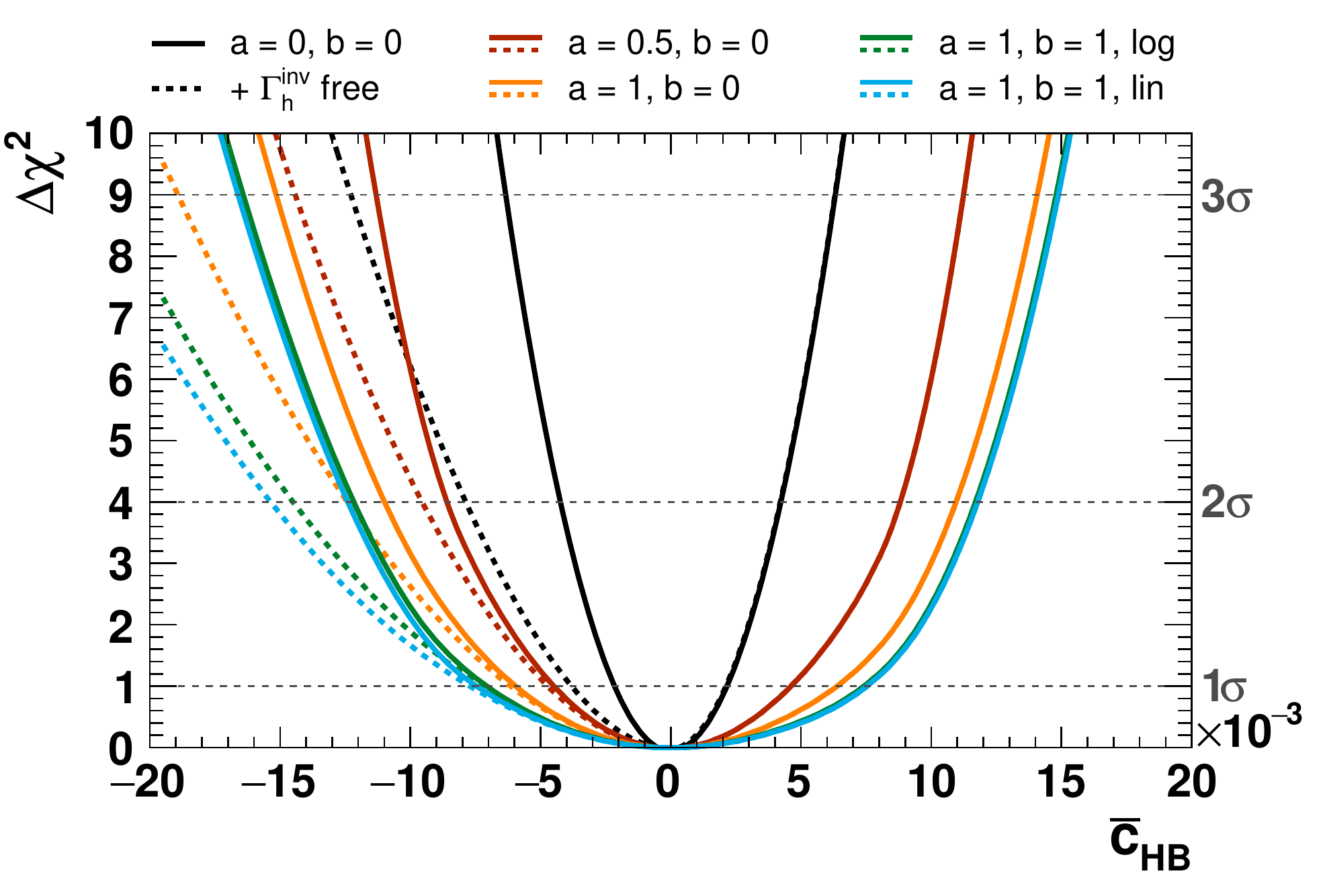}
	\hfill
	\includegraphics[width=0.47\textwidth]{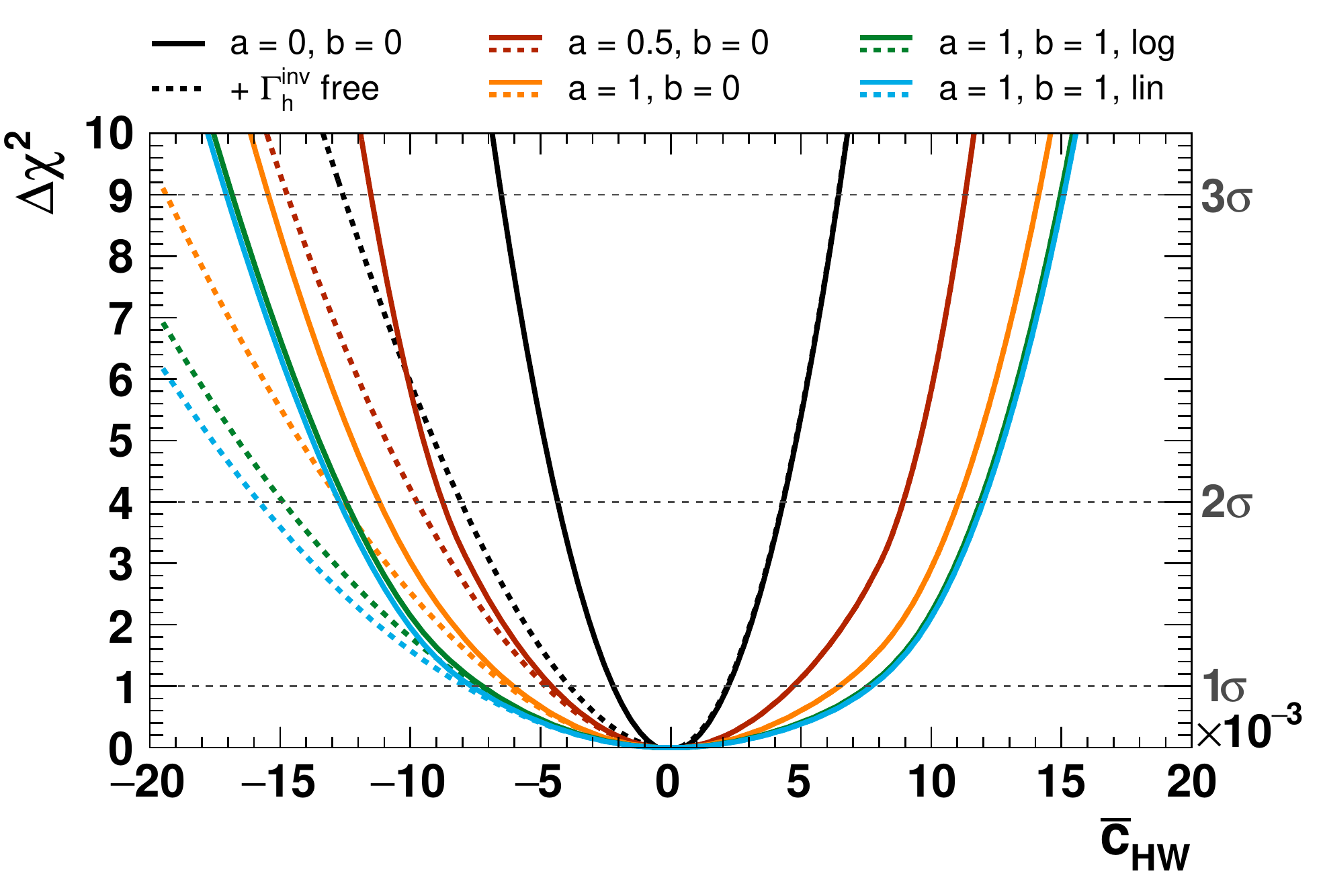} \\
	\includegraphics[width=0.47\textwidth]{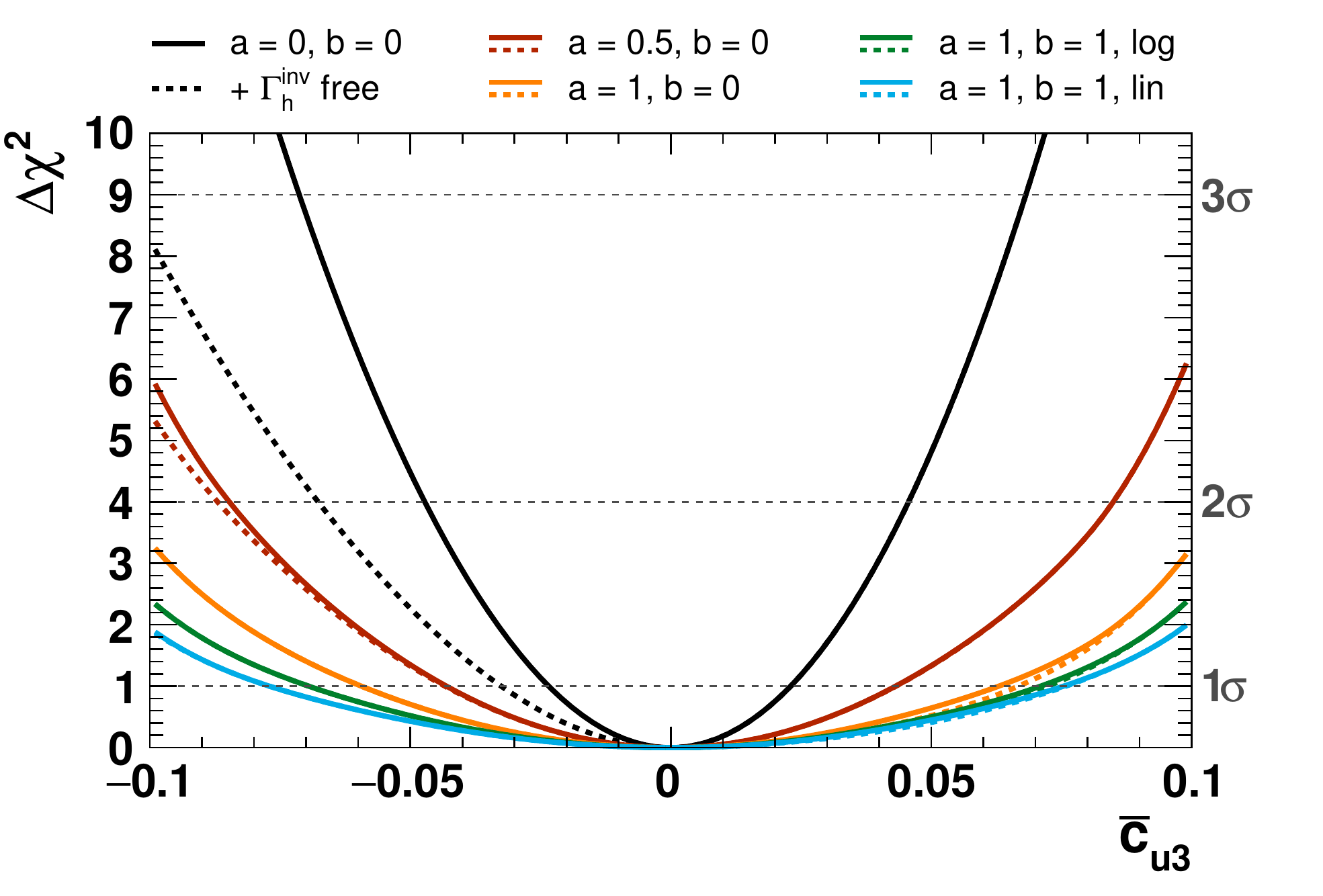}
	\hfill
	\includegraphics[width=0.47\textwidth]{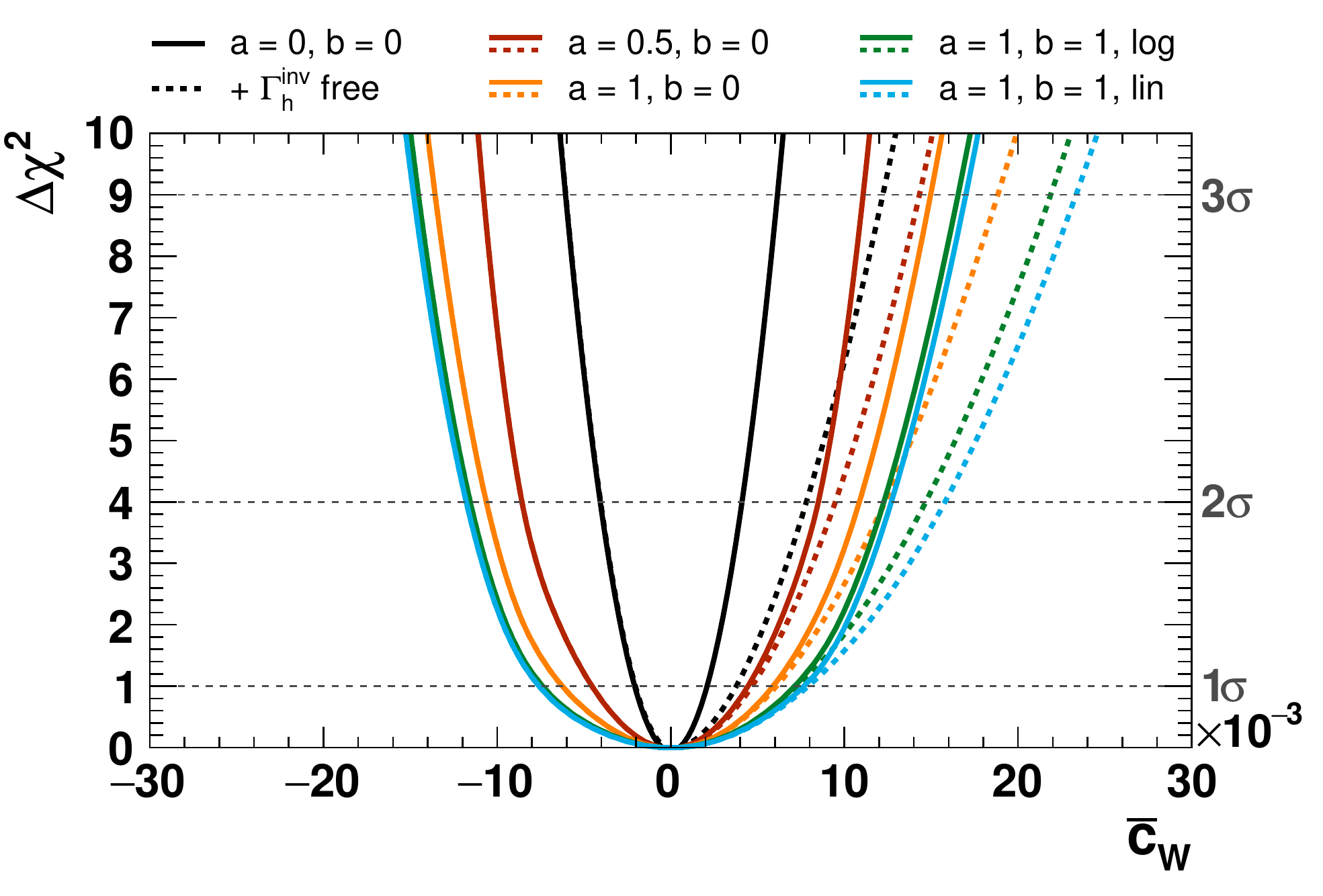} \\
	\caption{\label{fig:inc1}Scans of the sum of squared residuals $\Delta \chi^2$ as a function of the Wilson coefficients $\bar{c}_{d,3}$, $\bar{c}_{g}$, $\bar{c}_{\gamma}$, $\bar{c}_{H}$, $\bar{c}_{HB}$, $\bar{c}_{HW}$, $\bar{c}_{u,3}$, and $\bar{c}_{W}$ (from top left to bottom right), obtained by using differential Higgs $\pt$ distributions. Different assumptions on theoretical uncertainties as given in Eq.~\eqref{eq:uncertainties} are shown in different colours. Solid lines refer to scenarios with $\Gamma_h^{\text{inv}}=0$, while dashed lines indicate fit results with $\Gamma_h^{\text{inv}}$ left free in the fit. }
\end{figure*}
%%%%%%%%%%%%%%%%%%%%

%%%%%%%%%%%%%%%%%%%%
\begin{figure*}[!t]
	\subfigure[]{\includegraphics[width=0.47\textwidth]{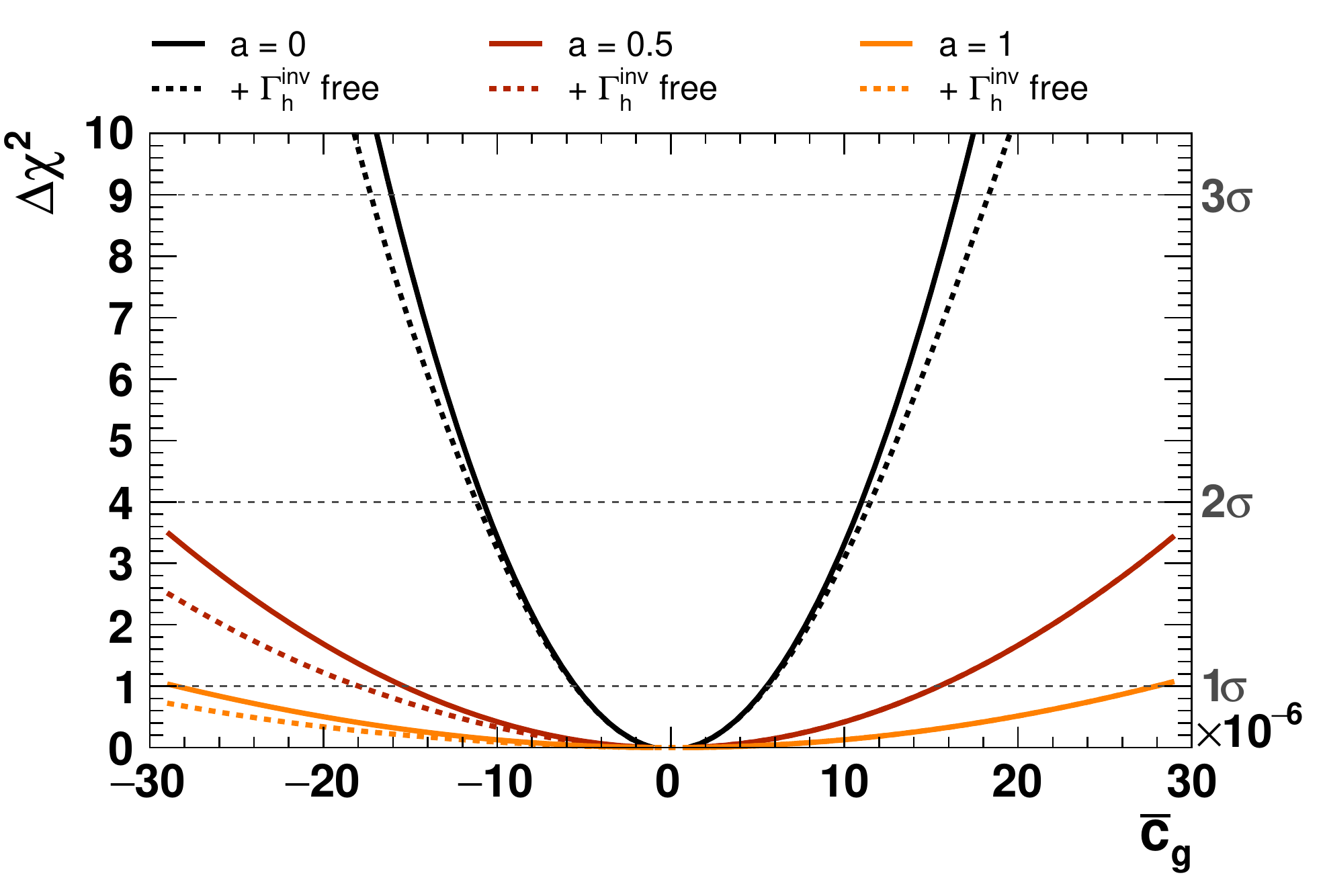}}
	\hfill	
	\subfigure[]{\includegraphics[width=0.47\textwidth]{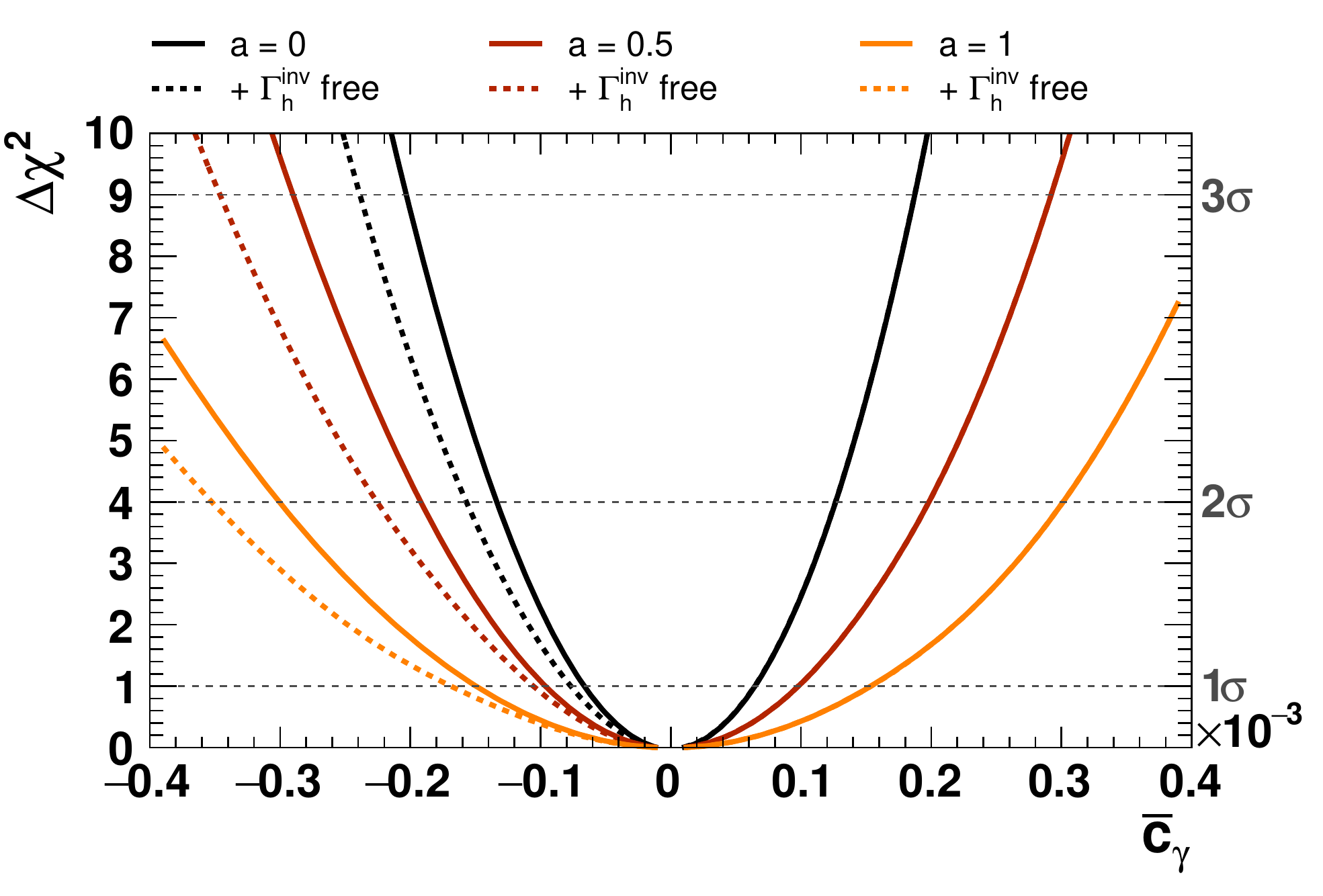}} 
	\subfigure[]{\includegraphics[width=0.47\textwidth]{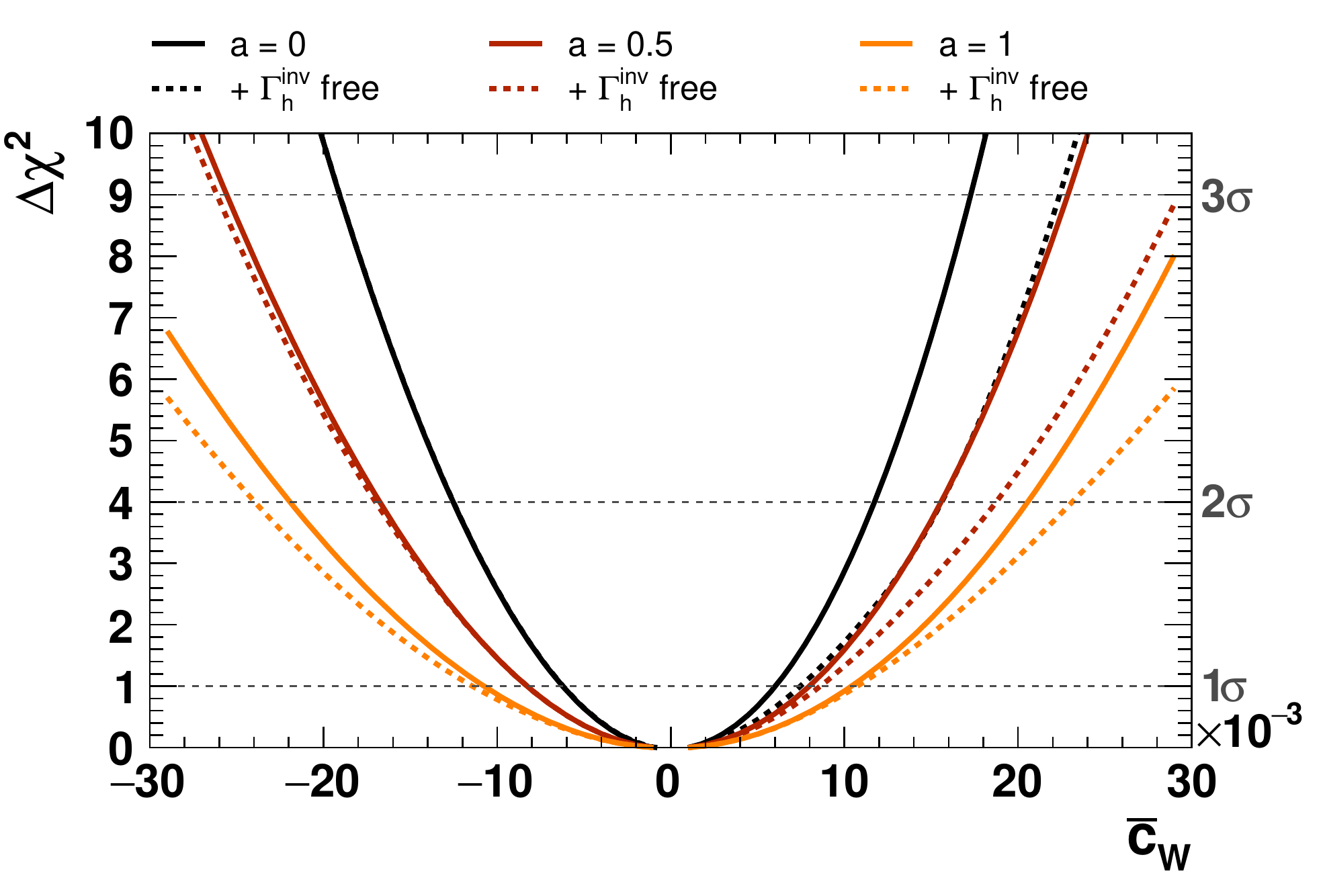}}
	\hfill	
	\subfigure[]{\includegraphics[width=0.47\textwidth]{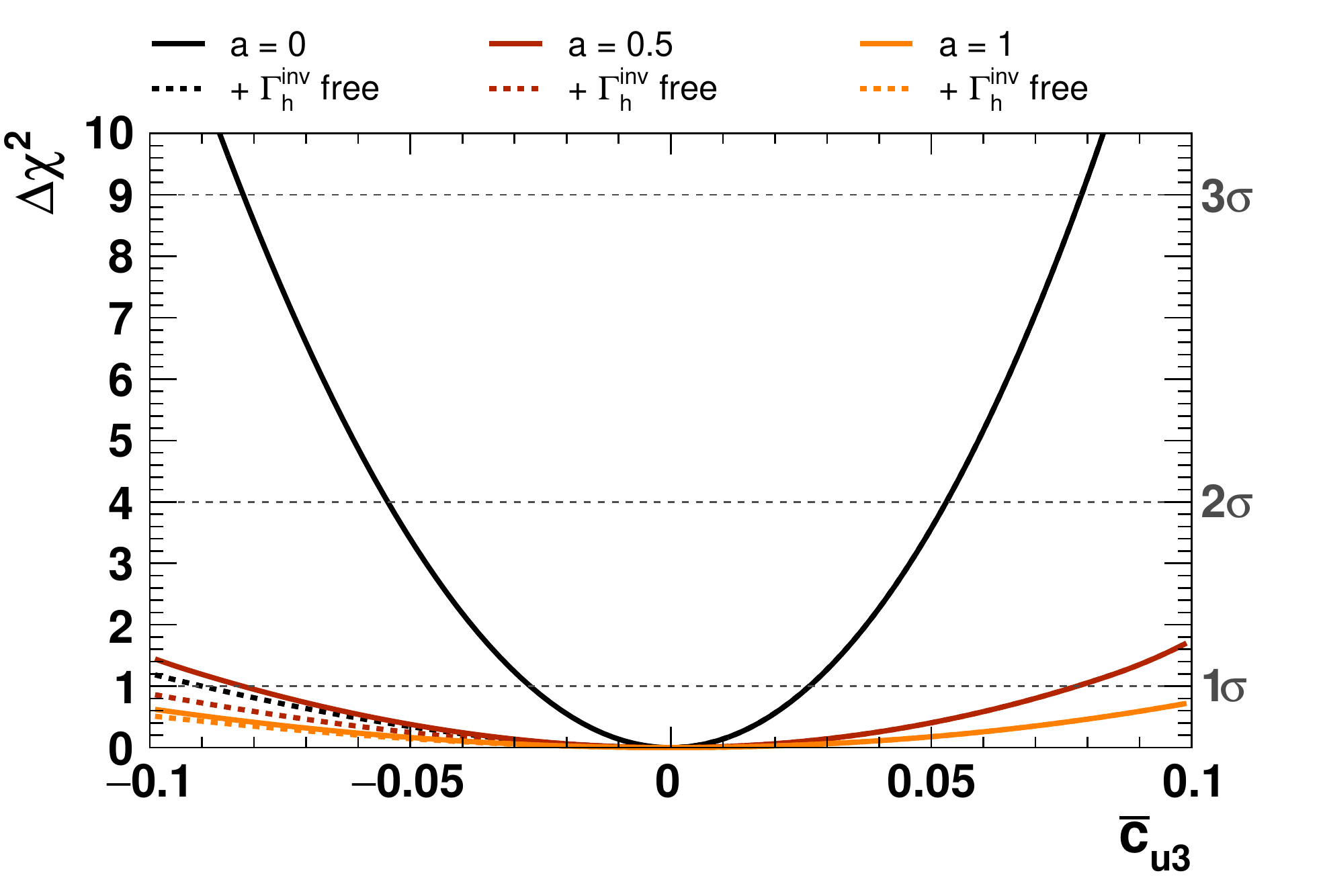}} 
	\caption{\label{fig:inc_sig}Scans of the sum of squared residuals $\Delta \chi^2$ as a function of the Wilson coefficients $\bar{c}_{g}$, $\bar{c}_{\gamma}$, $\bar{c}_{W}$ and $\bar{c}_{u,3}$ (from top left to bottom right), obtained by using signal strength measurements only. Different assumptions on theoretical uncertainties as given in Eq.~\eqref{eq:uncertainties} are shown in different colours. Solid lines refer to scenarios with $\Gamma_h^{\text{inv}}=0$, while dashed lines indicate fit results with $\Gamma_h^{\text{inv}}$ left free in the fit.  }
\end{figure*}
%%%%%%%%%%%%%%%%%%%%

A crucial question when analysing the impact of differential distributions on the Higgs characterisation programme is the level at which theoretical as well as experimental uncertainties can limit the sensitivity. When setting constraints on continuum deviations from the SM, the larger theoretical uncertainties that are intrinsic to the perturbative modelling of large momentum-transfer final states compared to inclusive quantities significantly degrade the sensitivity to relative excesses, as expected from Eq.~\eqref{eq:silh}. Hence the naive expectation that the overflow bins (or the very large transverse momentum regions) provide the largest statistical pull in a fit is typically misleading for Wilson coefficient choices that warrant the use of perturbative techniques~\cite{Englert:2014cva,Englert:2016aei,Contino:2016jqw}.\footnote{Note, however, that a non-perturbative Wilson coefficient constraint remains a physical statement as the validity of the parameter range is only gauged by matching the EFT to a concrete UV scenario.} In practice, the most sensitive region in a fit is given by the region of phase space where BSM deviations are large compared to theoretical as well as experimental uncertainties. However, as we will see below, the importance of the tails of distributions also depends on the concrete physics question that we would like to investigate.

A practical problem then arises when trying to provide sensitivity estimates for a large statistical sample of expected LHC data.\footnote{Current LHC measurements, which constrain the Higgs couplings at the 10\% level, are just about providing a larger sensitivity to BSM-induced modifications than expected from electroweak precision constraints in selected scenarios.} In the following we will focus (and compare) a range of parametrisations of theoretical uncertainties in the Higgs transverse momentum distribution $\ptH$ (this observable is likely to be reported in unfolded form~\cite{deFlorian:2016spz}) and trace their impact through the fit procedure. Concretely, we choose a functional form of the theory uncertainty of 
\begin{subequations}
\label{eq:uncertainties}
\begin{equation}
	\delta(\ptH) = \delta_0 [a + b f(\ptH)]\,.
\end{equation}
The parameter $\delta_0$ refers to the inclusive cross section uncertainties. We employ two parameterisations for the $\ptH$-dependence
\begin{alignat}{5}
\text{(i)}  & \quad f(\ptH) = \log \left(1 +{\ptH \over m_H} \right) \,,\\
\text{(ii)} & \quad f(\ptH) = {\ptH \over m_H} \,.
\end{alignat}
\end{subequations}
A linear scaling of the theoretical uncertainties is undoubtedly a very conservative outlook into the future while a logarithmic scaling is motivated from QCD considerations~\cite{Caola:2016upw}. The two terms in Eq.~\eqref{eq:uncertainties}, corresponding to an uncertainty in the inclusive cross section ($\sim a$) and an uncertainty in the tails of the \ptH distributions ($\sim b$), are allowed to vary independently in the fits.

In Fig.~\ref{fig:inc1}, we show constraints obtained from $\ptH$ distributions for the uncertainty choices detailed above. These constraints document a categorisation of Wilson coefficients that explicitly distinguish between the sensitivity in the Higgs decay or its production, or in both decay and production. For example, not being able to experimentally resolve $b\bar b H$ production, the sensitivity to $\bar{c}_{d,3}$ comes exclusively from branching ratio modifications, while $\bar{c}_g$ impacts both, in particular the most dominant gluon fusion production mechanism. 

This categorisation (within our approximations, for the off-shell $gg\to ZZ$ constraints see below) pinpoints to the sensitivity gain that can be reached from improved theory uncertainties.
The results of Fig.~\ref{fig:inc1} already provide a telltale story of the relative statistical power of high momentum transfer final states. On the one hand, operators like $\sim \bar{c}_{g}$ (as well as all other operators which induce a momentum dependence in Higgs production) will sculpt the differential distribution. On the other hand, operators like $\sim \bar{c}_{d,3}$ (or all other light flavour quark and lepton operators), which have a suppressed contribution to the production phenomenology and predominantly impact the decay of the Higgs, will globally shift the distribution and are therefore only constrained by the absolute cross section measurement (or signal strength) and the associated uncertainty. As a consequence $\bar{c}_{d,3}$ is not impacted too much by how we model the expected uncertainty in the tails of the Higgs $\pt$ distributions but is saturated by the total fiducial uncertainty of the combination of Higgs measurements. This is different for operators $\sim \bar{c}_{g}$ which change the shape of the distributions towards harder or softer $\ptH$ spectra (depending on the Wilson coefficients' sign) at a given observed signal strength. Here, the tail uncertainties play a crucial role in constraining the new physics-induced functional deviation of the Higgs distributions in the light of the theoretical and experimental uncertainties.

Starting from an idealised base-line where we assume all theoretical uncertainties to be absent (black line) in Fig.~\ref{fig:inc1}, the biggest relative deterioration of the limits we find when including a flat uncertainty band across all bins of the $\ptH$ differential distribution (red and orange lines). The orange lines correspond to a band of size of the current uncertainties on the inclusive Higgs production processes (see also \cite{Englert:2015hrx}), and the red line assumes a 50\% improvement over time. Inflating the uncertainties in the tails further, using either a linear or logarithmic function, has only a minor effect on all operators, with the marked exception of $\bar{c}_g$.

We contrast the constraints from differential distributions of Fig.~\ref{fig:inc1} with signal-strength-only measurements in Fig.~\ref{fig:inc_sig}. Limits derived from signal-strength measurements are entirely based on the total number of reconstructed Higgs boson events in each final state. Theoretical uncertainties only enter via coefficient $a$ in Eq.~\eqref{eq:uncertainties}, whereas $b=0$ throughout. They are particularly suited to set limits within the so-called $\kappa$-framework \cite{LHCHiggsCrossSectionWorkingGroup:2012nn}, a much simpler theoretical framework compared to the SM EFT approach, where a scale-dependence of production and decay rates is absent, and both are thus rescaled globally without phase-space dependence. However, within the SM EFT framework differential distributions provide crucial information to set much tighter limits on the effective operators at hand, see also Ref.~\cite{Englert:2015hrx}. For the operators shown in Fig.~\ref{fig:inc_sig} we find consistently a significant improvement using differential distributions, compared to signal strength measurements, with the only exception of $\bar{c}_g$ and $\bar{c}_{u_3}$ in the absence of theoretical uncertainties. While absence of any theoretical uncertainty is an unrealistic scenario, it shows that $\bar{c}_g$ and $\bar{c}_{u_3}$, both contributing directly to the dominant Higgs production mechanism at the LHC, receive strong constraints from Higgs threshold production. On the one hand, this observation motivates the precise calculation of the process $gg \to H$~\cite{Anastasiou:2016cez}, while on the other hand one can deduce that a greater effect can be achieved by reducing theoretical uncertainties in the tail of the $H+\mathrm{jet}$~\cite{Boughezal:2015dra} and  $H+2~\mathrm{jet}$~\cite{Campbell:2012am} distributions. For those operators the utilisation of differential distributions not only constrains them individually, but also allows one to resolve the blind direction in parameter space $\bar{c}_g + \bar{c}_{u_3} \simeq 0$~\cite{Harlander:2013oja,Grojean:2013nya, Schlaffer:2014osa,Buschmann:2014twa,Dawson:2014ora,Dawson:2015gka}.  

%%%%%%%%%%%%%%%%%%%%
\begin{figure*}[!t]
	\subfigure[]{\includegraphics[width=0.48\textwidth]{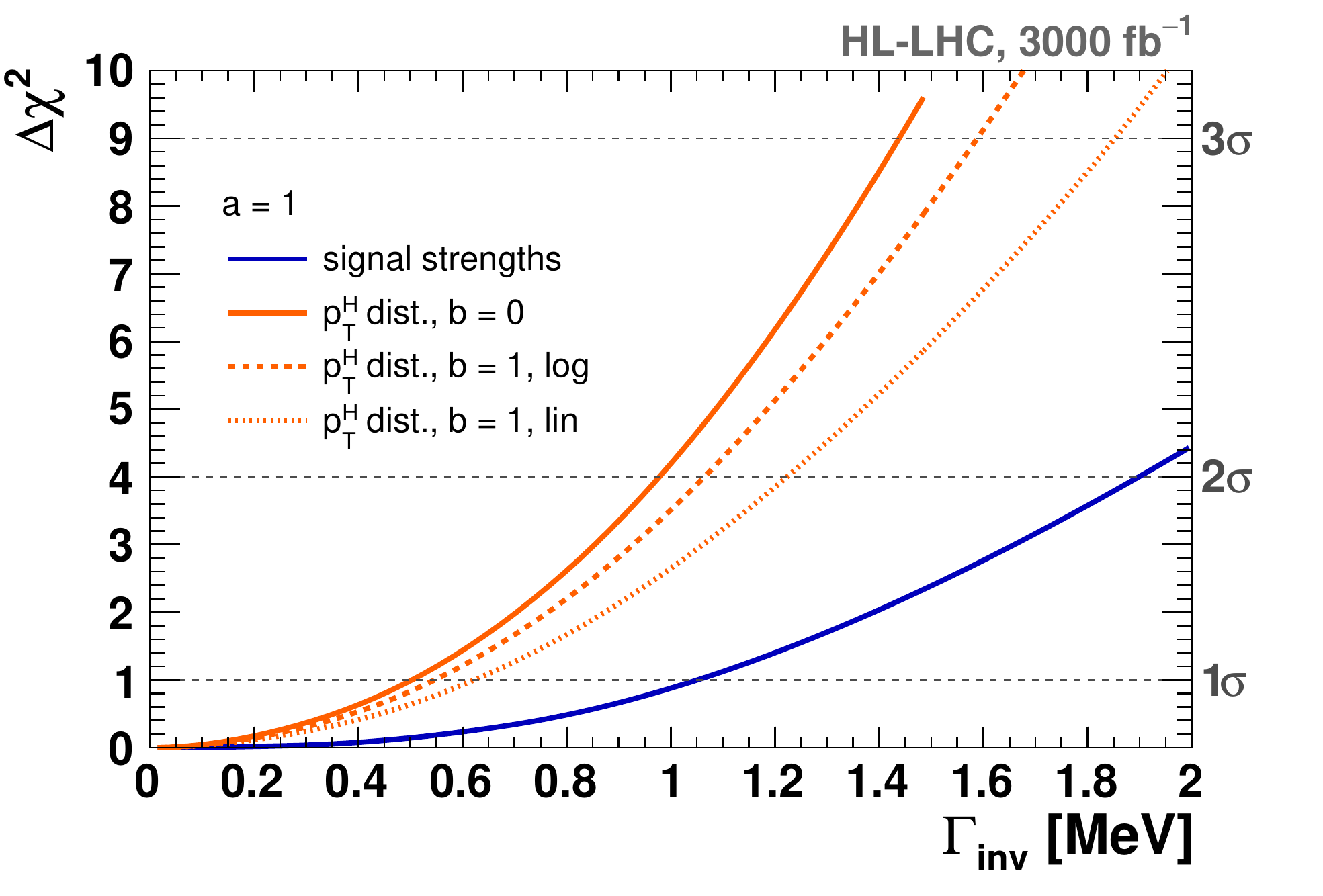}}
	\hfill	
	\subfigure[]{\includegraphics[width=0.48\textwidth]{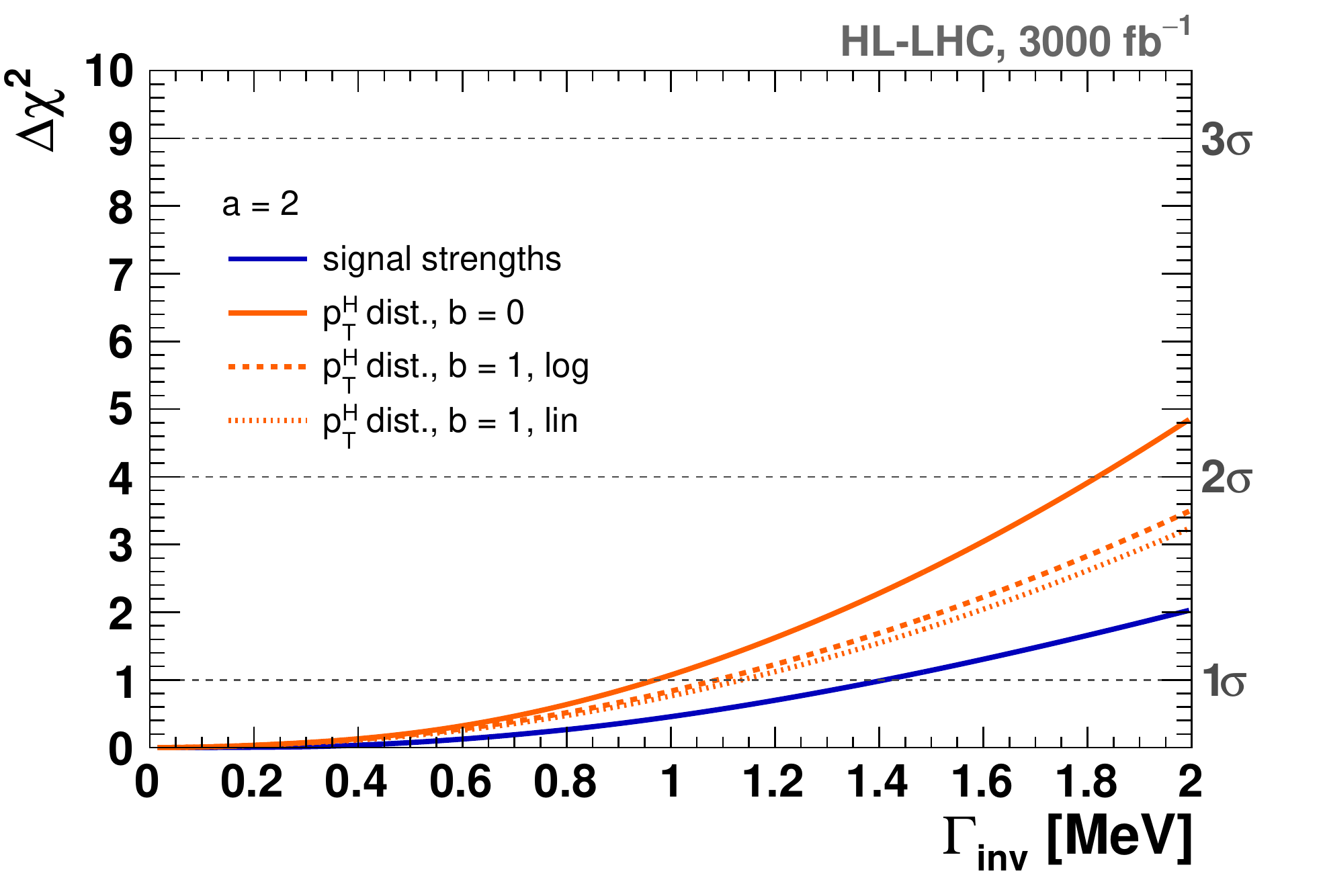}}
	\caption{Constraints on the total width extraction $\Gamma_h^{\text{inv}}$ through signal strength measurements only (blue line) and through differential $\ptH$ measurements (orange lines), for the different choices of $\ptH$-dependent uncertainties. \label{fig:hwidth1}}
\end{figure*}
%%%%%%%%%%%%%%%%%%%%

Extending the Higgs SM EFT Lagrangian of Eq.~(\ref{eq:silh}) by a light degree of freedom to allow for invisible or undetectable Higgs decays (dashed lines in Figs.~\ref{fig:inc1} and \ref{fig:inc_sig}), according to Eq.~(\ref{eq:modhiggs}), has only a small effect on the operators proportional to $\bar{c}_{d,3}$, $\bar{c}_g$, $\bar{c}_\gamma$, $\bar{c}_H$ and $\bar{c}_{u,3}$. However, the subset of operators mediating interactions between the Higgs and electroweak gauge bosons, i.e.\ operators proportional to $\bar{c}_{HB}$, $\bar{c}_{HW}$ and $\bar{c}_{W}$, show a much bigger dependence on the presence of such a decay mode. The reason for this is the following. The total width $\Gamma_h$ is mostly constrained by the decays $H \to b\bar{b}$ and $H\to WW$, where $H \to b\bar{b}$ is dominantly controlled through $\bar{c}_{d,3}$, but $H\to WW$ receives contributions from a larger set of operators. The operators proportional to $\bar{c}_{HB}$, $\bar{c}_{HW}$ and $\bar{c}_{W}$ can compensate for changes in $H\to WW$, and therefore for changes in $\Gamma_h$, when varying one of these three Wilson coefficients at a time, leading to a weaker constraint on $\Gamma_h^{\text{inv}}$ and consequently on the operator itself. In contrast to this, when varying a coefficient like $\bar{c}_{\gamma}$, the total width $\Gamma_h$, and therefore also $\Gamma_h^{\text{inv}}$, are constrained sufficiently through channels not affected by changes in $\bar{c}_{\gamma}$, and the constraints on the corresponding operator are only very weakly affected by invisible or undetectable Higgs decays.

%%%%%%%%%%%%%%%%%%%%
\section{Constraining the total Higgs width}
\label{sec:conshiggs}
From our discussion of the previous section, we can now already anticipate the results for the constraints from differential Higgs kinematics on the extraction of the total Higgs width in the singlet-extended scenario of Eq.~\eqref{eq:modhiggs}. Indeed, as shown in Fig.~\ref{fig:hwidth1}, the tail uncertainties do not impact the extraction of the Higgs width too significantly and the inclusive measurements provide the strongest sensitivity. The limits obtained on the invisible branching ratio from this fit procedure are in the 20\% range for the chosen form of theoretical uncertainties, $a=1$ in Eq.~\eqref{eq:uncertainties}. This result, however, is extremely sensitive to the inclusive uncertainties. Doubling the theoretical uncertainties by a factor of 2, $a=2$, the limit on the invisible branching ratio deteriorates to $\sim 30\%$. Constraints in the vicinity of $\sim 10\%$, which are favoured by scenarios of dark matter-related portal scenarios~\cite{Djouadi:2011aa,Lebedev:2011iq, Craig:2014lda} seem to be unattainable from Higgs measurements alone. Note that we have not included dedicated analyses for hidden Higgs decays. Available analyses by ATLAS and CMS~\cite{Aad:2015txa,Khachatryan:2016whc} place the branching ratio constraint in the vicinity of 20\%, for SM-like Higgs production (reducing to $\sim 40\%$ for non-SM production~\cite{Khachatryan:2016whc}). As this is already comparable to the extraction of the Higgs width from visible on-shell data in the EFT context, we can expect these channels to also play a dominant role in constraining invisible Higgs decays beyond the limitations set by visible channels. Cross section ratios of tagged production categories in hidden Higgs decay analyses against their experimentally observable counterparts will provide additional constraints from a plethora of search channels.

%%%%%%%%%%%%%%%%%%%%
\subsection{Relevance of the ``off-shell'' measurement \label{sec:offshell}}

%%%%%%%%%%%%%%%%%%%%
\begin{figure*}[!t]
	\subfigure[]{\includegraphics[width=0.48\textwidth]{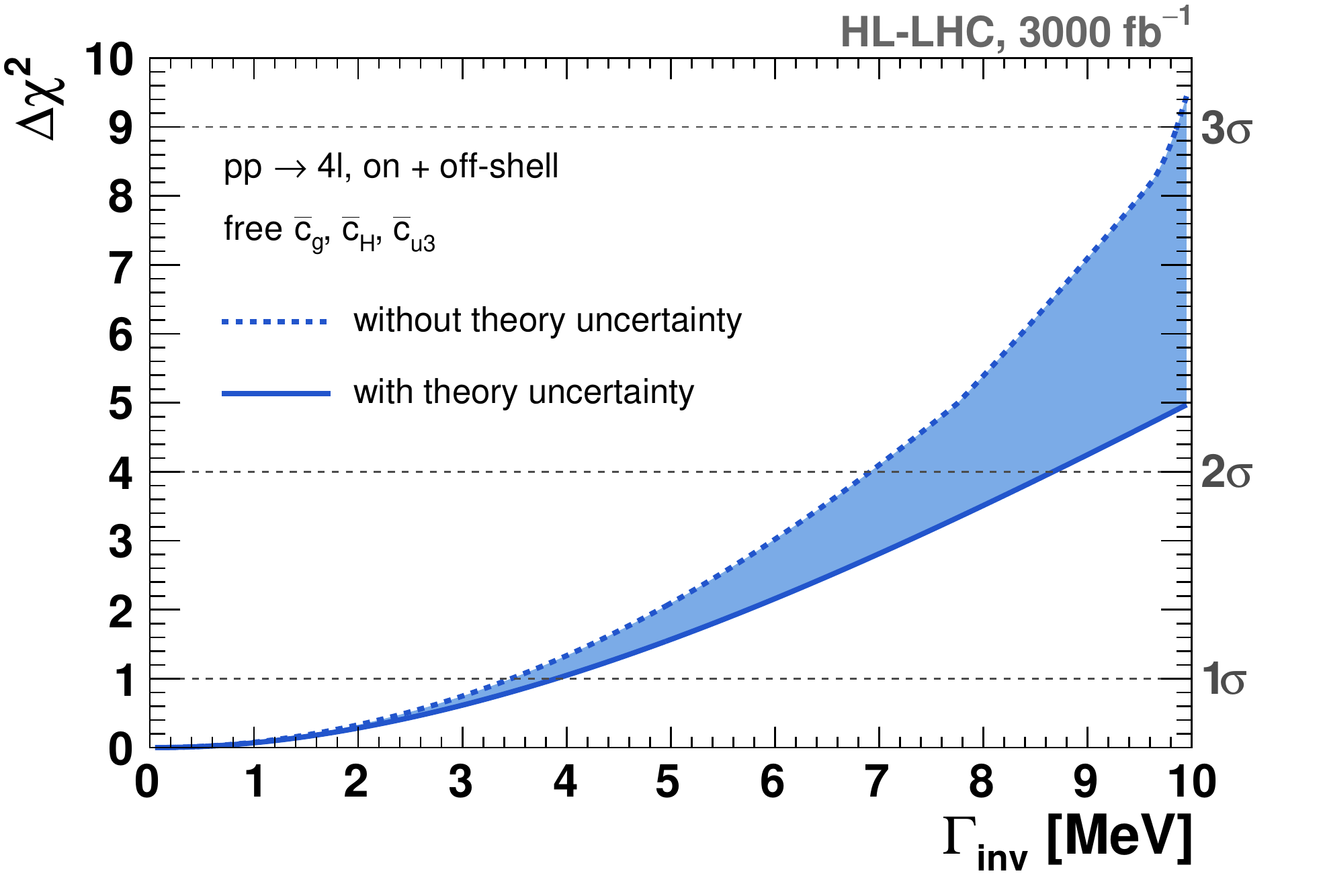}}
	\hfill	
	\subfigure[]{\includegraphics[width=0.48\textwidth]{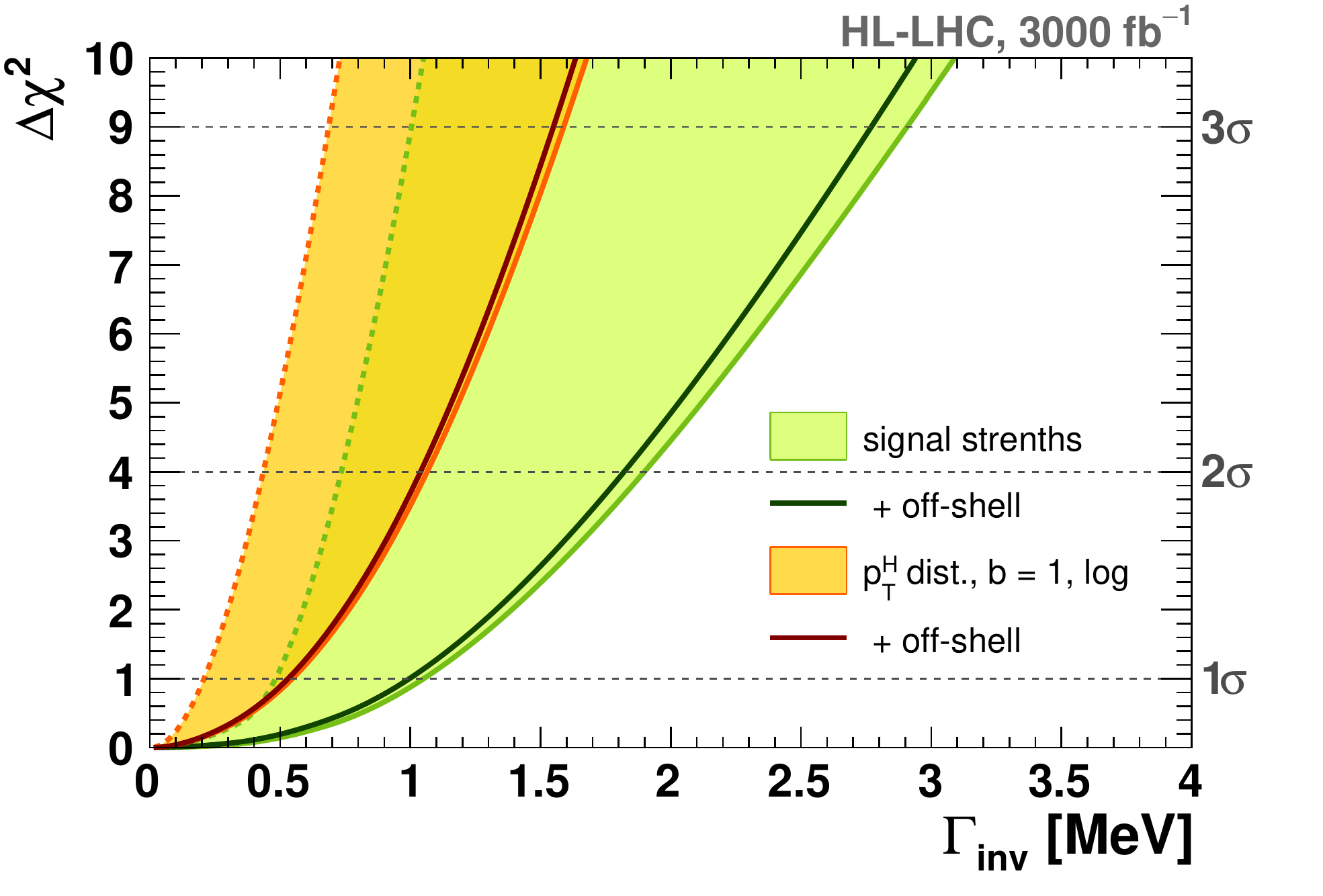}}
	\caption{Higgs width extraction from the off-shell measurement of $gg\to ZZ$. (a) Limiting the fit to $\bar{c}_g, \bar{c}_H, \bar{c}_{u3}$ and (b) the overall impact of the off-shell measurement in comparison with the on-shell measurements. The choice of $a=1$ is made for the theoretical uncertainties and their impact is depicted as coloured bands.\label{fig:hwidth2}}
\end{figure*}
%%%%%%%%%%%%%%%%%%%%

We come to the question if the so-called ``off-shell measurement'' in $gg\to ZZ$ final states that has gained importance as a probe of the Higgs width in $\kappa$-like interpretations of Higgs coupling measurements provides an additional handle in constraining the Higgs lifetime~\cite{Caola:2013yja} (see in particular Ref.~\cite{Corbett:2015ksa} for fitting results along these lines). The assumption of a simple coupling-modifier correlated modification of on-shell $gg\to H\to ZZ$ to off-shell $gg\to ZZ$ phase-space regions allows one to interpret an overproduction of the $ZZ$ tail in terms of larger width suppression of an on-shell observation of the Higgs boson. Turning this argument around, any cross section limit away from resonance in the $ZZ$ channel can be interpreted as an upper limit on the Higgs width, baring in mind the imposed correlation of the different phase-space regions~\cite{Englert:2014aca,Englert:2014ffa}. The $gg\to ZZ$ channel is particularly appealing as the continuum probes perturbative unitarity cancellations of massive fermion scattering $t\bar t \to ZZ$ above the Higgs threshold $m_{ZZ}$ through the absorptive parts of the $gg\to ZZ$ amplitude in a phenomenologically clean way, which invalidates the naive zero width approximation and implies a non-decoupling behaviour of the continuum~\cite{Kauer:2012hd,Kauer:2013cga} (see also Ref.~\cite{Campbell:2013una}).

Through opening up the correlations between on-shell and off-shell region as induced by the complexity of Eqs.~\eqref{eq:silh} and~\eqref{eq:modhiggs}, the specific assumptions of the $\kappa$ coupling modifier do in general not hold under the more general assumptions of the fit and an SM-like off-shell measurement does not necessarily imply a SM-like coupling structure. As on-shell and off-shell region become largely decorrelated through the number of competing new interactions, the on-shell measurement acts as the driving force of the Higgs width measurement as the off-shell region (given by the experimental selection $m_{ZZ}>330~\mathrm{GeV}$) has lost all memory of the Higgs width by construction. The off-shell measurement hence constitutes in the EFT extension of Eq.~\eqref{eq:silh} only an additional measurement, whose sensitivity to the modified Higgs diagrams is only weak~\cite{Kauer:2012hd}. The comparably large statistical uncertainty expected in the off-shell region can be expected to result in only a mild constraint.

This expectation is validated by Fig.~\ref{fig:hwidth2}. We see only a mild improvement of the invisible width constraint compared to the signal-strength and $\ptH$-driven constraints. The on-shell measurements leave too much freedom in the extended coupling space for the off-shell measurement to be relevant in the light of the expected statistical uncertainty.

%%%%%%%%%%%%%%%%%%%%%%%%%%%%%%%%
\section{Summary and conclusions}
\label{sec:conc}
In this work we have assessed the impact of theoretical uncertainties on generic searches for new physics beyond the SM in the Higgs sector through differential distributions and connected these findings to resonant invisible extensions of SM EFT, where the Higgs obtains an hidden-sector partial decay width. Since our analysis does not include any information on invisible Higgs searches, our results are also applicable to invisible Higgs decay modes, e.g.\ $H\to \text{light flavour quarks}$.

While the Higgs width is not a free parameter in SM EFT but fixed through the choice of SM couplings, masses, and Wilson coefficients, the extension discussed in this work is a minimal simplified model that allows one to address the question of the Higgs lifetime in a theoretically controlled way. We find that a hidden branching ratio of $\sim 10\%$ seems accessible at the high luminosity LHC. Theoretical uncertainties remain limiting factors for such a search.

A measurement that has been highlighted as a sensitive probe of the Higgs width in the literature is the off-shell measurement in $gg\to ZZ$. Including this channel to our fit, we only find a small added sensitivity compared to on-shell measurements alone. The reason behind this is that, while the tail of the distribution carries some information about the Wilson-coefficients off-shell, the decay properties are determined by on-shell measurements, which can be sufficiently accessed by inclusive Higgs measurements across the standard search channels. This behaviour has been anticipated before, and our results directly validate previously made arguments~\cite{Englert:2014ffa}.

\acknowledgments 
This research was supported in part by the German Research Foundation (DFG) in the Collaborative Research Centre (SFB) 676 ``Particles, Strings and the Early Universe'' located in Hamburg. MS is supported in part by the European Commission through the "HiggsTools" Initial Training Network PITN-GA- 2012-316704.

%%%%%%%%%%%%%%%%%%%%%%%%%%%%%%%%
\bibliography{references}
%%%%%%%%%%%%%%%%%%%%%%%%%%%%%%%%

\end{document}